\documentclass{iaus}

\usepackage{graphicx}

\title[Symp. 262.~~Stellar Populations - Planning for the Next Decade] 
{What does the IMF really tell us\\ about star formation?}

\author[Kouwenhoven \& Goodwin]
{M.B.N. Kouwenhoven$^{1,2}$ \and S.P. Goodwin$^2$}

\affiliation{
$^1$The Kavli Institute for
Astronomy and Astrophysics, Peking University, Yi He Yuan Lu 5, Hai
Dian Qu Beijing 100871, P.\,R. China \\ email: {\tt kouwenhoven@kiaa.pku.edu.cn} \\
[\affilskip]
$^2$University of Sheffield, Hicks Building, Hounsfield
Road, Sheffield S3\,7RH, United Kingdom \\email: {\tt s.goodwin@sheffield.ac.uk} \\
}

\pubyear{2009}
\volume{262}  
\pagerange{1--2}
\jname{S262 Stellar Populations - Planning for the Next Decade}
\editors{G. Bruzual et al., eds.}
\begin{document}

\maketitle

\begin{abstract}
Obtaining accurate measurements of the initial mass function (IMF) is
often considered to be the key to understanding star formation, and a
universal IMF is often assumed to imply a universal star formation
process. Here, we illustrate that different modes of star formation
can result in the same IMF, and that, in order to truly understand
star formation, a deeper understanding of the primordial binary
population is necessary. Detailed knowledge on the binary fraction,
mass ratio distribution, and other binary parameters, as a function of
mass, is a requirement for recovering the star formation process from
stellar population measurements.  
\end{abstract}


\noindent Finding the IMF has been a major goal of stellar and
galactic astrophysics, as it provides crucial information about the
star formation process (e.g., Salpeter 1955; Kroupa 2002; Chabrier
2003; Bonnell, Larson \& Zinnecker 2007).  The core mass function
(CMF) was shown to have a shape very similar to the IMF (e.g., Motte
et al. 2001; Simpson, Nutter \& Ward-Thompson 2008, and numerous
others), suggesting a link between the two (see, in particular, Alves,
Lombardi \& Lada 2007; Goodwin et al. 2008). However, this link is not
trivial.  It is known that most stars are part of a binary system
(Lada \& Lada 2003), and that most (possibly even all) stars have
formed in binary or multiple systems (e.g., Goodwin \& Kroupa 2005,
Kouwenhoven et al, 2005, 2007). The IMF represents the mass
distribution over all stars (including companion stars), and therefore
depends on the properties of the primordial binaries (in particular
the binary fraction and mass ratio distribution). The solution to the
problem of recovering the outcome of star formation process is not
unique, and the IMF therefore has to be interpreted with care.

\begin{figure}
  \centering
  \begin{tabular}{cccc}
    \includegraphics[width=0.19\textwidth]{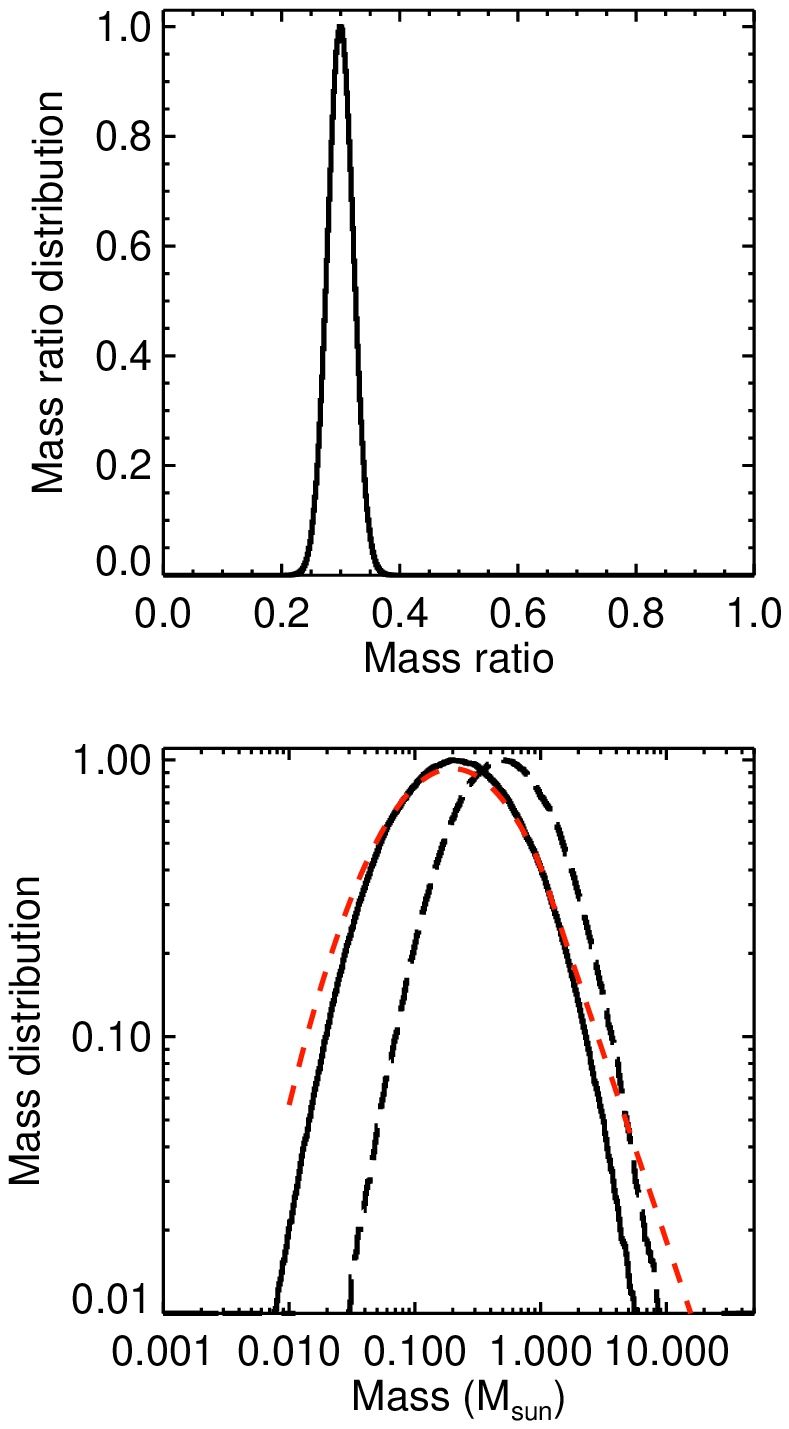} &
    \includegraphics[width=0.19\textwidth]{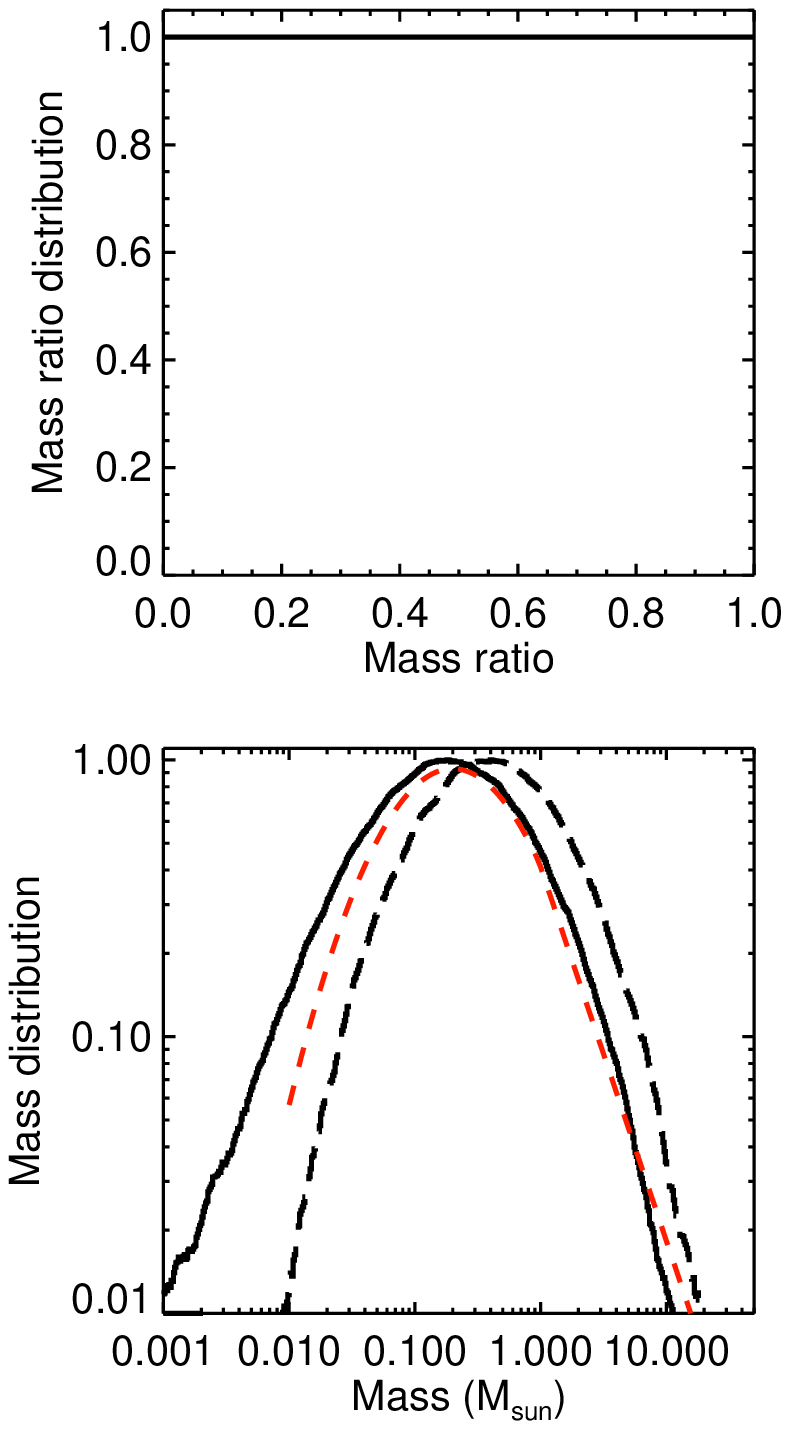} &
    \includegraphics[width=0.19\textwidth]{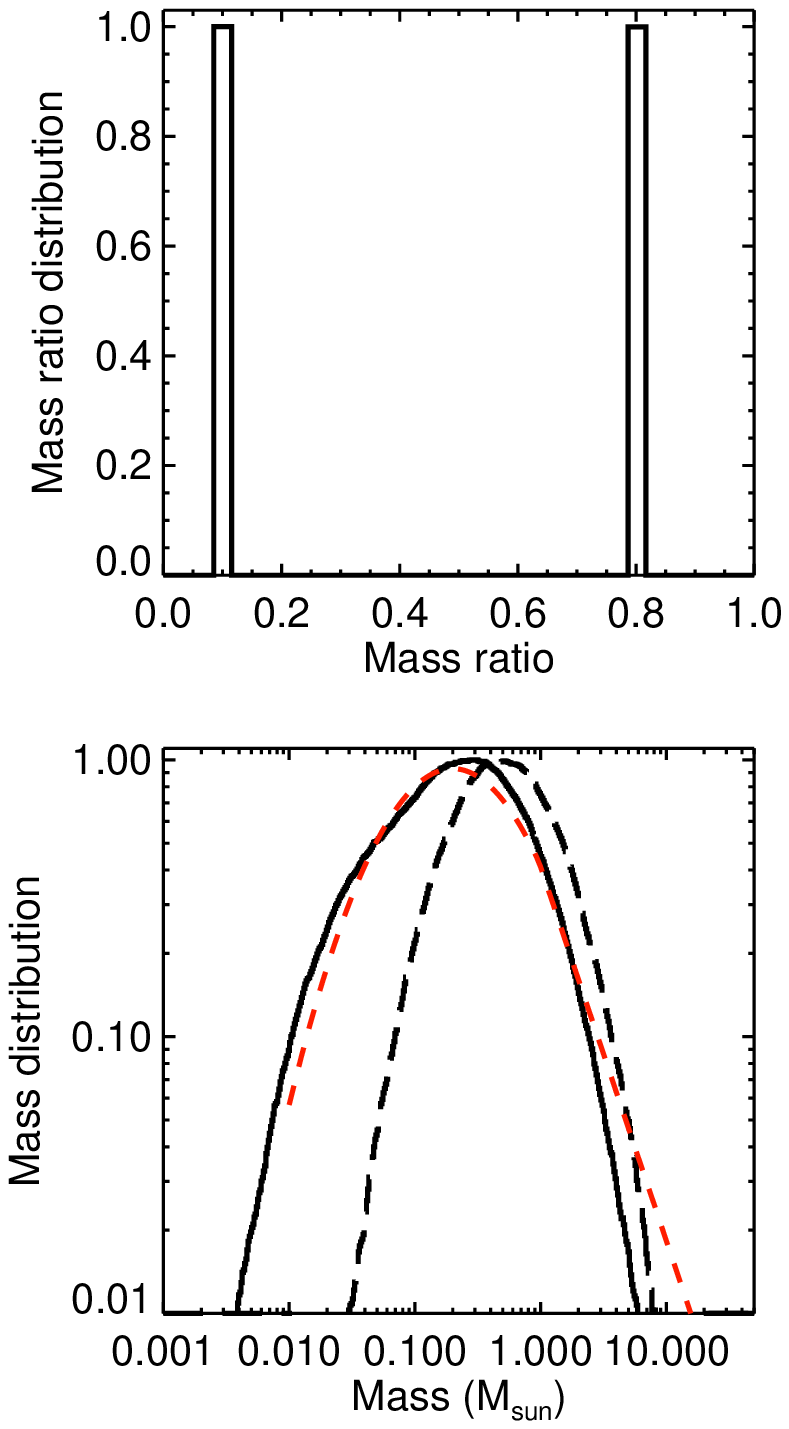} &
    \includegraphics[width=0.19\textwidth]{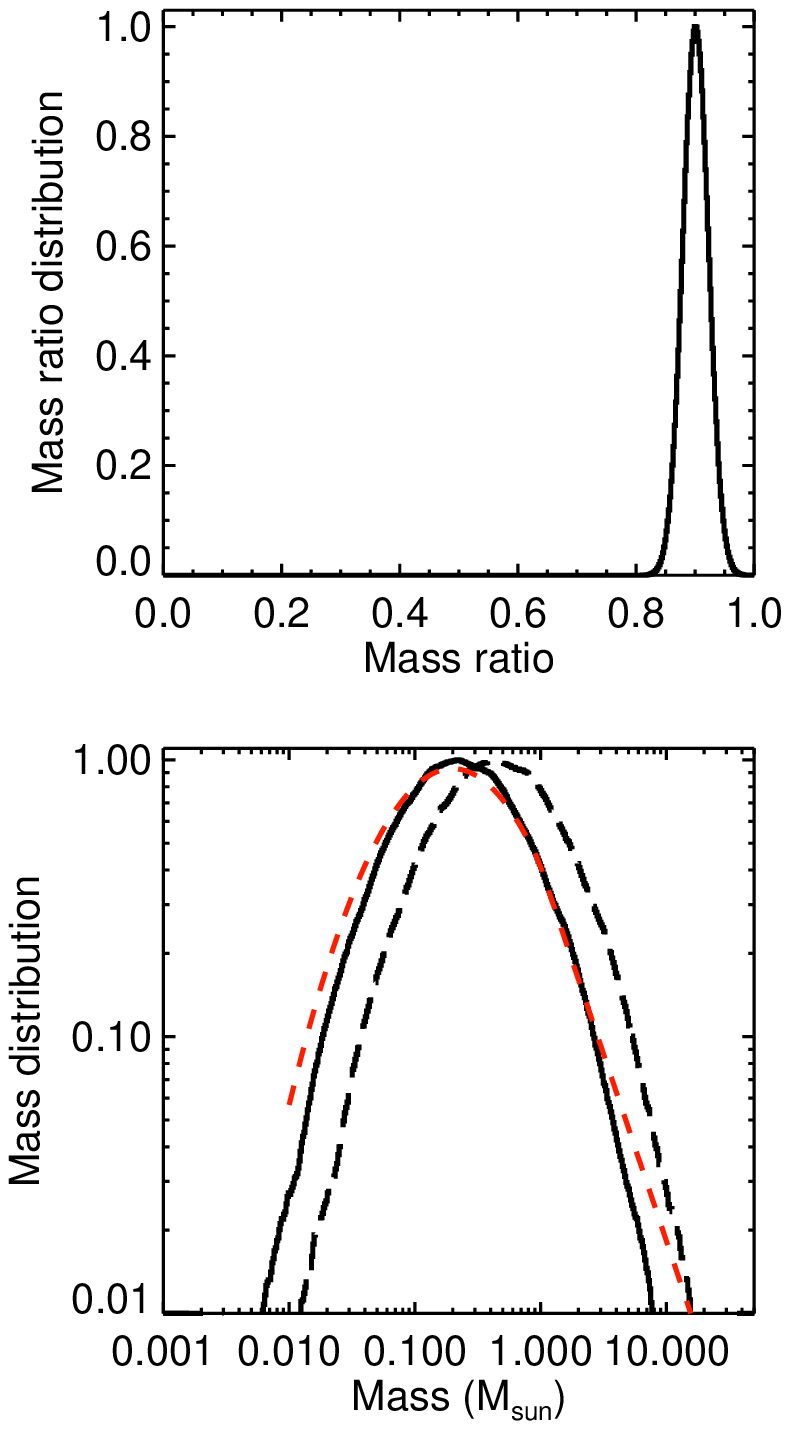} \\
  \end{tabular}
  \caption{In the bottom panels we show the IMFs (solid black lines)
    formed from a given system mass function (SMF; dashed line) with an
    $f(q)$ as shown in the top panel, assuming a binary fraction of
    unity. The IMFs are compared to a canonical Chabrier (2003) IMF
    (light red dashed lines). The means and variances of the SMFs that
    give the best fits are (from left to right) $\mu_{\log 10 M} =
    -0.3$, $-0.4$, $-0.3$ and $-0.35$, and $\sigma_{\log 10 M}=0.4$,
    $0.55$, $0.4$, and $0.5$, respectively.  }
\end{figure}

To illustrate this issue, we generate stellar populations from a CMF,
and compare the IMFs resulting from different formation processes
(i.e., different binary fractions and mass ratio distributions).  We
assume a universal CMF similar the observed CMF of Alves et al. (2007)
and Nutter \& Ward-Thompson (2007), which has an average mass of $\sim
1 M_\odot$, and a width of $\sigma_{\log M_C} \sim 0.5$.  Each core is
then converted into a system mass $M_S$, given a certain core-to-star
formation efficiency (SFE). Finally, a fraction of these systems is
split up into binaries with a certain mass ratio distribution $f(q)$,
following the prescription for SCP-I described in Kouwenhoven et
al. (2009).

Fig.~1 shows the IMFs (bottom panels) resulting from models with
different mass ratio distributions (top panels), where we have set the
binary fraction and SFE to unity. Despite the star formation process
(here represented by the mass ratio distribution) being completely
different, the IMFs are similar. In fact, the four IMFs are
statistically identical for any observational sample with a realistic
number of stars. The results can be generalised for populations with a
mass-dependent binary fraction, $f(q)$, and SFE; for details we refer
to Goodwin \& Kouwenhoven (2009).  With the examples shown in Fig.1 we
do not imply that all modes of star formation result in the same IMF,
but merely that it is not possible to constrain star formation using
the IMF alone.

It is often assumed that stellar populations with the same IMF must
have had an identical formation process. Here, we illustrate that
completely different modes of star formation may result in the same
IMF, and that an observed IMF has to be interpreted with care. On the
other hand, if two star clusters are found to have a different IMF,
star formation must have been different. 
The IMF is an extremely important concept in understanding star
formation, and theories and simulations of star formation should
result in an IMF identical to the observed IMF. The general shape of
the IMF is primarily determined by the properties of the CMF. However,
to recover the details of star formation, the IMF alone is not enough;
measuring the binary fraction, $f(q)$, and other binary parameters as
a function of stellar mass is a prerequisite.


\vskip2mm \noindent {\bf Acknowledgements}. MBNK was supported by the
Peter and Patricia Gruber Foundation through the PPGF fellowship, and
by PPARC/STFC (grant number PP/D002036/1).

\end{document}